\documentclass{article}
\usepackage[utf8]{inputenc}

\usepackage{amsmath, amsthm, amssymb}
\usepackage{mathabx}
\usepackage{mathtools}
\usepackage{graphicx}
\usepackage{verbatim}
\usepackage{natbib}
\usepackage{color} 
\usepackage{caption}
\usepackage{subcaption}
\usepackage{fancyvrb}
\usepackage{enumerate}
\usepackage{relsize}
\usepackage{xcolor}

\usepackage{float}

\usepackage{hyperref}
\usepackage[margin=1.2in]{geometry}
\hypersetup{colorlinks,citecolor=blue,urlcolor=blue,linkcolor=blue}

\usepackage[shortlabels]{enumitem}
\usepackage{algorithm2e}
\RestyleAlgo{boxruled}
\LinesNumbered
\DontPrintSemicolon
\SetAlFnt{\normalsize}
\SetKwInOut{Input}{Input}

\usepackage{nikos_tex}

\theoremstyle{plain}

\theoremstyle{definition}

\theoremstyle{remark}

\date{October, 2024}
\title{\vspace{-2cm}Empirical Bayes estimation via data fission}

\author{ Nikolaos Ignatiadis\\
  \texttt{ignat@uchicago.edu} \\
  \and 
  Dennis L. Sun \\
  \texttt{dlsun@stanford.edu} 
}

\begin{document}

\maketitle

\begin{abstract}
\noindent \textbf{Abstract:} We demonstrate how data fission, a method for creating synthetic replicates from single observations, can be applied to empirical Bayes estimation. This extends recent work on empirical Bayes with multiple replicates to the classical single-replicate setting. The key insight is that after data fission, empirical Bayes estimation can be cast as a general regression problem.\\

\noindent \footnotesize{This note was prepared as a comment on ``Data Fission: Splitting a Single Data Point,'' by James Leiner, Boyan Duan, Larry Wasserman, and Aaditya Ramdas, a discussion paper in the \emph{Journal of the American Statistical Association}.}
\end{abstract}

\vspace{0.8cm}


We congratulate Leiner, Duan, Wasserman, and Ramdas on 
a stimulating article that joins an 
elegant method to a compelling application. 
Their article focuses primarily on applications
to selective inference. In this comment, we 
demonstrate how data fission can be applied to a 
very different problem: empirical Bayes (EB) estimation~\citep{robbins1956empirical, efron2019bayes}. 

In the EB framework, we observe $X_1, \dotsc, X_n$ generated by 
\begin{equation}
\label{eq:basic_eb}
\theta_i \simiid H,\;\;\; X_i \mid \theta_i \simindep p(\cdot \mid \theta_i).
\end{equation}
If the prior $H$ were known, then 
the Bayes estimator $\hat\theta_i^B = \EE[H]{\theta_i | X_i}$ would be optimal, achieving the Bayes risk. In EB, the prior $H$ is not known, so the goal 
is to construct an estimator $\hat{\theta}^{EB}_i$ that approximates $\hat{\theta}^{B}_i$ using all of $X_1,\dotsc,X_n$.
\citet{ignatiadis2023empiricala} demonstrated 
a general way to construct EB estimators 
when i.i.d.\ replicates $X_{i1}, \dotsc, X_{iK}$ are 
available for each $\theta_i$. This method, called Aurora, regresses one replicate on the rest.

\begin{figure}
    \centering
    \begin{tabular}{l}
    (a) $X_i \mid \theta_i \sim \text{Normal}(\theta_i, 1)$ \\ 
    \includegraphics[width=1.1\linewidth]{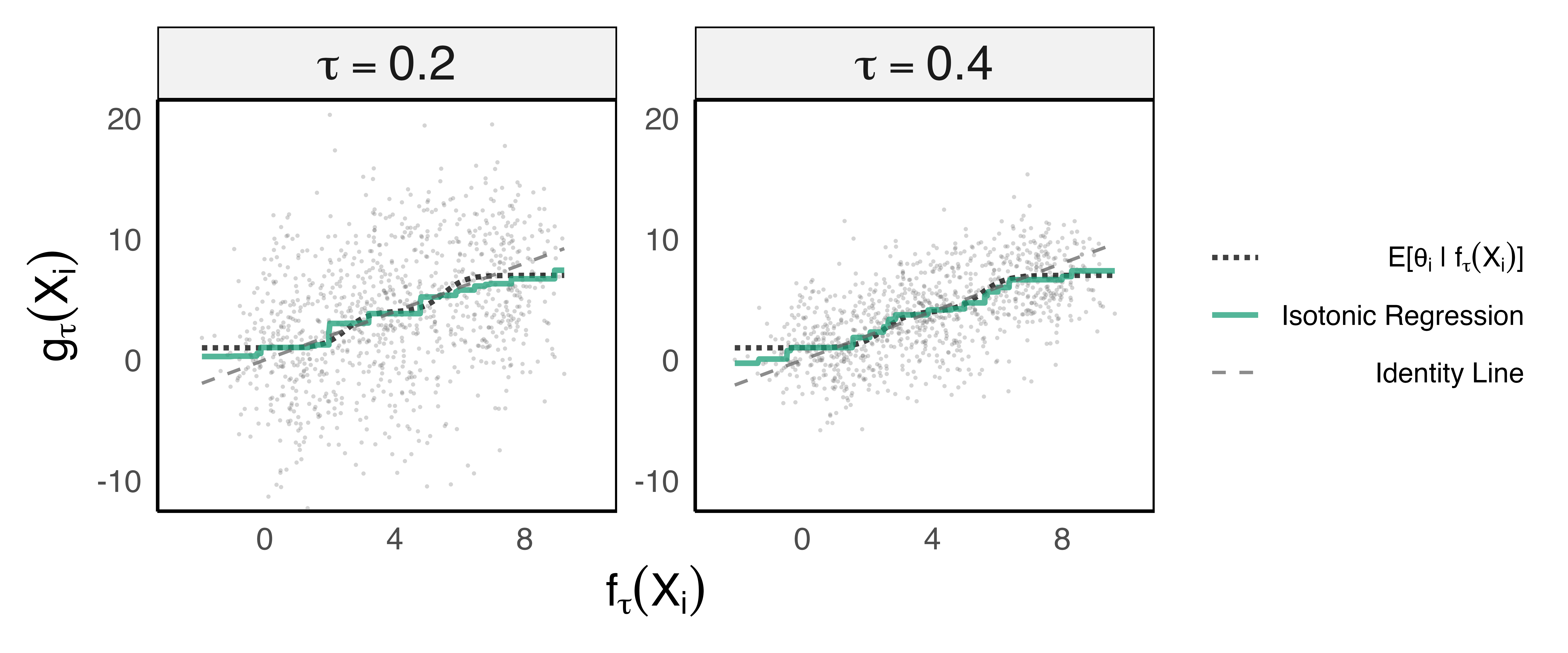} \\ 
    (b) $X_i \mid \theta_i \sim \text{Poisson}(\theta_i)$ \\ 
    \includegraphics[width=1.1\linewidth]{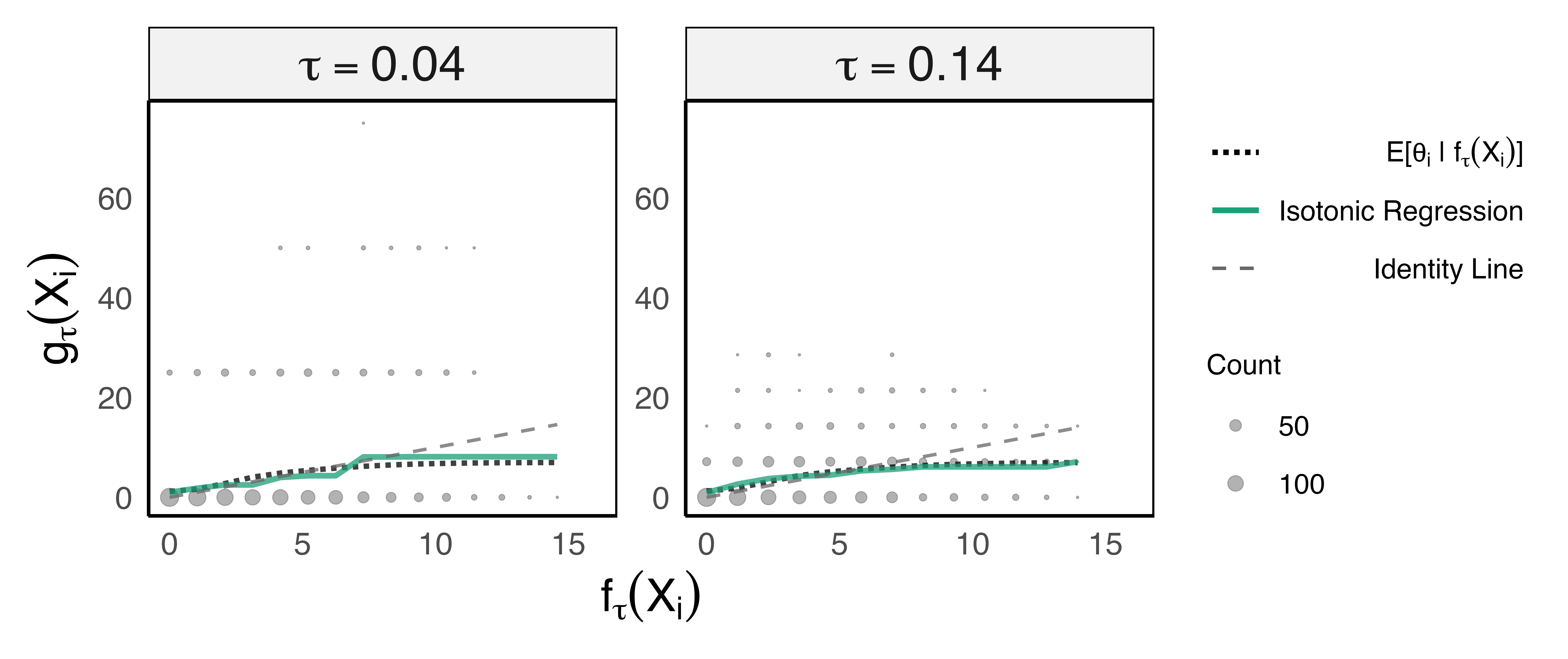} \\ 
    \end{tabular}
    \caption{We simulated $n=1000$ observations, with $\theta_i$ from the three-point prior $H=\mathrm{Unif}\cb{1,4,7}$, 
    and $X_i \mid \theta_i$ from either a normal or a Poisson
    distribution. Each $X_i$ was split into two 
    replicates, $f_\tau(X_i)$ and $g_\tau(X_i)$, for two values 
    of $\tau$. Each panel shows a scatterplot of the replicates
    for a different value of $\tau$ and a different likelihood, 
    along 
    with the true and estimated mean functions (estimated by isotonic regression). For smaller $\tau$, $f_{\tau}(X_i)$ contains more information about $\theta_i$, so the oracle estimator $\EE{\theta_i \mid f_{\tau}(X_i)}$ is more similar to the Bayes estimator $\EE{\theta_i \mid X_i}$. However, $g_{\tau}(X_i)$ is noisier, which makes the regression task more difficult. Following~\citet{leiner2023data}, we may interpret the values of $\tau$ for the normal and Poisson simulations as the split of the Fisher information between $g_{\tau}(X_i)$ and $f_{\tau}(X_i)$: in the left panels (small $\tau$), we have a split of 4:96 and in the right panels (medium $\tau$) a split of 14:86.  
    }
    \label{fig:normal_simulation}
\end{figure}

But what if there is only one $X_i$ per $\theta_i$?
This is where data fission comes in. 
We can use data fission to generate synthetic 
replicates and apply Aurora:

\begin{enumerate}
\item As in~\citet{leiner2023data}, we 
construct functions $f_\tau$ and $g_\tau$ appropriate to the likelihood $p(\cdot \mid \theta_i)$,  
with the EB-specific requirement that  $\EE[H]{g_{\tau}(X_i) \mid f_{\tau}(X_i)} = \EE[H]{\theta_i \mid f_{\tau}(X_i)}$.
    \begin{itemize}
    \item For a normal likelihood (with variance $\sigma^2$), the construction in 
    \citet{leiner2023data},
    $f_{\tau}(X_i) = X_i + \tau Z_i$ and $g_{\tau}(X_i) = X_i - \tau^{-1}Z_i$ with independent $Z_i \sim \text{Normal}(0, \sigma^2)$, works.
    \item For a Poisson likelihood, we rescale
    the construction in~\citet{leiner2023data} (also used in the EB context by~\citet{brown2013poisson}). Let \smash{$Z_i \simindep \mathrm{Bin}(X_{i}, 1-\tau)$} for $\tau \in (0,1)$, and set $f_{\tau}(X_i) =  Z_i/(1-\tau)$ and $g_{\tau}(X_i) =  (X_i-Z_i)/\tau$.
    \end{itemize}

\item Regress $g_{\tau}(X_i)$ on $f_{\tau}(X_i)$, $i=1,\dotsc,n$ using any regression method. Denote the estimated mean function by $\hat{m}(\cdot)$. 
\item Estimate $\theta_i$ by $\hat{\theta}_i = \hat{m}(f_{\tau}(X_i))$.  
\item Repeat data fission multiple times and average the resulting 
estimates.
\end{enumerate}

If the mean function $\hat{m}(\cdot)$ is learned well, then the risk of Aurora should approximately match the risk of $\EE[H]{\theta_i \mid f_{\tau}(X_i)}$. This expectation 
can be made arbitrarily close to the Bayes estimator 
by choosing $\tau$ small so that $f_\tau(X_i) \approx X_i$.
However, the variance of 
$g_\tau(X_i)$ also increases for small $\tau$, 
making the mean function harder to learn.

Figure~\ref{fig:normal_simulation} illustrates Aurora 
on simulated data. The scatterplots show replicates generated
by data fission for two values of $\tau$ and for 
normal and Poisson likelihoods. 
The true mean functions $\EE{g_\tau(X_i) \mid f_\tau(X_i)}$ are graphed as dotted black lines. The mean functions are estimated by isotonic regression and the estimates $\hat m(\cdot)$
are graphed as solid teal lines.

Table~\ref{tab:combined_simulation_results} compares the MSE of Aurora with the MLE ($\hat{\theta}_i = X_i$), the nonparametric maximum likelihood estimator (NPMLE), and the oracle Bayes estimator. The NPMLE performs best on these well-specified low-dimensional EB problems, as is known in the literature \citep{jiang2009general, koenker2014convex, polyanskiy2021sharp}. Yet, Aurora remains competitive, developing a classical connection between EB and regression~\citep{stigler19901988} into a general 
methodology (see~\citet{jana2023empirical, barbehenn2023nonparametric} for related ideas). Because Aurora is built on top of regression, 
it generalizes naturally to situations with 
side-information~\citep{ignatiadis2019covariatepowered} and high-dimensional likelihoods~\citep{daras2023ambient}.
And unlike the NPMLE, Aurora works even when $p(\cdot \mid \theta_i)$ is not fully specified, as long as data fission or data splitting is possible.

\paragraph{Reproducibility.} We provide code to reproduce our numerical results on Github:\\
\url{https://github.com/nignatiadis/empirical-bayes-data-fission-comment}

\begin{table}
    \centering
    \begin{tabular}{lcc}
        \hline
        Method & Gaussian MSE & Poisson MSE \\
        \hline
        MLE ($\hat{\theta}_i = X_i$) & 1.00 & 4.01 \\
        NPMLE (via \texttt{REBayes},~\citealp{koenker2017rebayes}) & 0.60 & 2.02 \\
        \hline
        Aurora with Isotonic Regression (small $\tau$) & 0.69 & 2.17 \\
        Aurora with Isotonic Regression (medium $\tau$) & 0.64 & 2.05 \\
        \hline
        Bayes Estimator (oracle, $\hat\theta_i^B = \EE[H]{\theta_i | X_i}$) & 0.59 & 1.97 \\
        \hline
    \end{tabular}
    \caption{Comparison of mean squared error (MSE) for different methods in the normal and Poisson simulations. The settings correspond to the panels of Figure~\ref{fig:normal_simulation}. We compute the MSE by averaging over $100$ Monte Carlo replicates of each simulation. We apply Aurora by averaging over $100$ repetitions of data fission. }
    \label{tab:combined_simulation_results}
\end{table}

\bibliographystyle{plainnat}
\bibliography{auroral}

\end{document}